\documentclass[pdflatex,sn-mathphys]{sn-jnl}

\jyear{2022}%

\theoremstyle{thmstyleone}%
\theoremstyle{thmstyletwo}%

\theoremstyle{thmstylethree}%

\begin{document}

\title[ ]{\textsc{Stereo} neutrino spectrum of $^{235}$U fission rejects sterile neutrino hypothesis}


\author{The \textsc{Stereo} collaboration\footnote{A list of authors and their affiliations appears at the end of the paper.}}

\abstract{Anomalies in past neutrino measurements have led to the discovery that these particles have non-zero mass and oscillate between their three flavors when they propagate. In the 2010's, similar anomalies observed in the antineutrino spectra emitted by nuclear reactors have triggered the hypothesis of the existence of a supplementary neutrino state that would be sterile i.e.{} not interacting via the weak interaction \cite{Mention:2011rk}. The \textsc{Stereo} experiment \cite{STEREO:2018blj,STEREOOscillationPhase1only,STEREO:2019ztb,STEREO:2020fvd,STEREO:2020hup} was designed to investigate this conjecture, which would potentially extend the Standard Model of Particle Physics.
Here we present an analysis of the full set of data generated by \textsc{Stereo}, confirming observed anomalies while rejecting the hypothesis of a light sterile neutrino.
Installed at the ILL (Institut Laue Langevin) research reactor, \textsc{Stereo} accurately measures the antineutrino energy spectrum associated to the fission of $^{235}$U. The segmentation of the detector and its very short distance to the compact core are crucial properties of \textsc{Stereo} for our analysis. The measured antineutrino energy spectrum suggests that anomalies originate from biases in the nuclear experimental data used for the predictions \cite{Mueller:2011nm,Huber:2011wv}. Our result supports the neutrino content of the Standard Model and establishes a new reference for the $^{235}$U antineutrino energy spectrum. We anticipate that this result will allow to progress towards finer tests of the fundamental properties of neutrinos but also to benchmark models and nuclear data of interest for reactor physics \cite{Estienne:2019ujo,Letourneau:2022kfs} and for observations of astrophysical or geo-neutrinos \cite{Lagage1985,Leyton2017}.
}




\maketitle

\section*{Introduction}\label{scn:introduction}

Reactor antineutrinos have played an important role in the history of neutrino physics contributing to the discovery of that particle in 1956 \cite{Cowan:1956rrn}. They are produced after the fission process during the beta-decay of fission fragments. Thus, intense antineutrino fluxes with a known flavor are produced by nuclear reactors, allowing to study precisely their properties. The increased accuracy in both prediction and  measurements has paved the way for the study of neutrino oscillations with essential results in the determination of mass splitting and mixing angles \cite{KamLAND:2008dgz,DoubleChooz:2019qbj,DayaBay:2016ggj,RENO:2018dro} that define the frequency and amplitude of such oscillations. A major advance in this field was the measurement of total beta spectra associated with the fission of $^{235}$U, $^{239}$Pu and $^{241}$Pu performed in the 1980s \cite{VonFeilitzsch:1982jw,Schreckenbach:1985ep,Hahn:1989zr}. The shape of these spectra was measured with a magnetic spectrometer \cite{MAMPE1978127} and the absolute normalisation determined by neutron capture reactions on gold and lead targets replacing the fissile isotopes for the calibration of the apparatus. A numerical procedure was then defined to convert these beta spectra into antineutrino spectra \cite{Vogel:2007du} which served as a reference for decades. In 2011, a re-evaluation of this procedure, called the ``Huber-Mueller'' model (HM) \cite{Mueller:2011nm,Huber:2011wv}, led to the ``Reactor Antineutrino Anomaly'' (RAA) in the comparison with data: the measured antineutrino rates were about 6\% below predictions on average \cite{Mention:2011rk}. A few years later a shape distortion was observed by experiments at power reactors in the form of a ``bump'' between 5 and 6 MeV with an amplitude of about 10\% \cite{DoubleChooz:2019qbj,DayaBay:2016ggj,RENO:2018dro,NEOS:2016wee}.
The disappearance of reactor antineutrinos as a function of their distance travelled from the reactor being the expected manifestation of an oscillation phenomenon, it was proposed that a new, fast oscillation towards an additional neutrino could explain the RAA deficit. From the existing constraints on the electroweak sector of the Standard Model \cite{ALEPH:2005ab}, this neutrino must be sterile, i.e. not coupling with the weak interaction; to be compatible with previous neutrino data its mass would be expected in the 1 eV range \cite{Mention:2011rk}.
The mixing of fission electron antineutrinos with a sterile neutrino state of such a mass can be approximately described by the two-flavor oscillation formula
\begin{equation}
\label{eq:oscillation}
        P_{ee} = 1 - \sin^2(2\theta_{ee}) \sin^2 \left(\frac{\Delta m^2_{41}L}{4E}\right)
\end{equation}
with $\sin^2(2\theta_{ee})$ the amplitude of the mixing, $\Delta m^2_{41}$ the square mass splitting between sterile and standard neutrino states, $L$ the distance between the emission and detection vertices and $E$ the energy of the antineutrino.

Sterile neutrino states are a common feature of models explaining the neutrino masses \cite{RevNuMass} or can also be dark matter candidates, triggering a worldwide experimental program to search for these particles \cite{Abazajian:2012ys}.

\section*{The \textsc{Stereo} experiment}\label{scn:experiment}

\begin{figure}[ht]
    \centering
    \includegraphics[width=\linewidth]{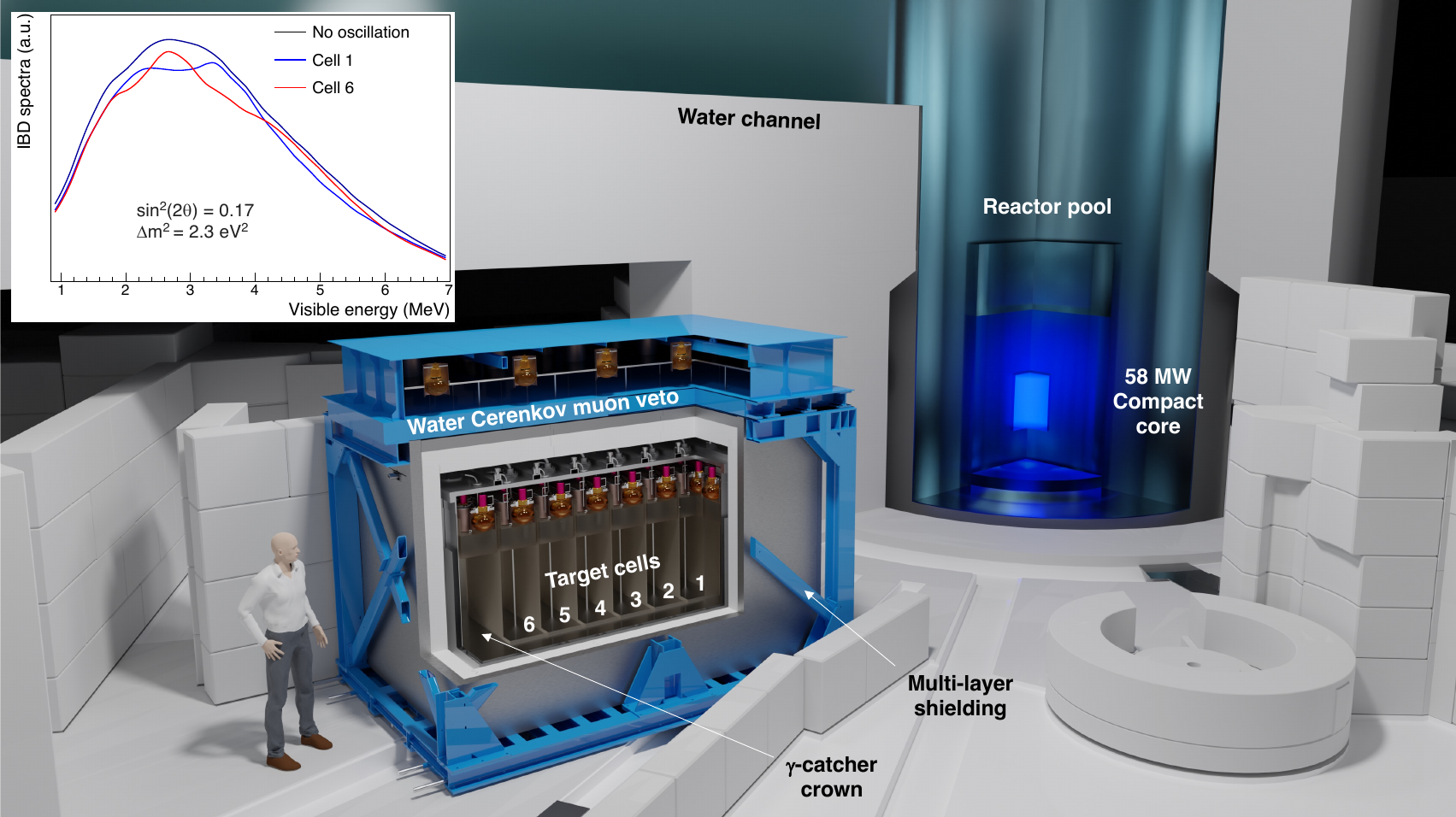}
    \caption{\textbf{Configuration of the STEREO experiment}. This drawing illustrates the proximity to the ILL compact core and the detector structure. The 6 identical cells of the target, numbered from 1 to 6 starting from the reactor side, are filled with 1.8 m$^3$ of gadolinium-doped liquid scintillator. Energy leakage from an antineutrino event close to a target edge is mitigated by the presence of a gamma-catcher surrounding the target volume and filled with Gd-unloaded liquid scintillator. 
    The walls are all reflective and improve the collection of scintillation light to photomultipliers tubes (PMT) located at the top of each cell. A 20 cm thick acrylic buffer separates the PMTs from the liquid to ensure a homogeneous response in the whole cell volume. A 1.5 mm layer of mu-metal protects the detector from external magnetic fields, and a polyethylene and lead shield isolates it from neutron and gamma fluxes. The whole structure is covered by a water Cerenkov detector allowing to veto cosmic muons passing near the target volume. A layer of soft iron and boron-loaded rubber completes the protection against magnetic fields and thermal neutrons respectively. Additional lead and polyethylene shielding at the front and sides of the detector is not shown for clarity. The presence of a water channel above the detector offers a crucial protection against the cosmic background by reducing the vertical muon flux by a factor of about 4 and by stopping the hadronic showers. The inset illustrates the spectral distortions expected in the first and the last cell in case of sterile-neutrino driven oscillations with parameters corresponding to the best fit of the RAA.
    }
    \label{fig:Stereo_Site}
\end{figure}

The \textsc{Stereo} detector was installed in 2016 at the ILL high-flux research reactor in Grenoble. From Eq.~\ref{eq:oscillation}, a coupling with an 1-eV sterile neutrino would result in spectral distortions over a few metres, as a function of the distance to the reactor core. This phenomenon is exploited by the segmentation of the target volume of \textsc{Stereo} into 6 identical cells (Fig.~\ref{fig:Stereo_Site}). The sterile neutrino hypothesis is thus tested by the comparison of the 6 measured spectra, independent of any prediction of the emitted spectrum.

Antineutrinos are detected via the inverse beta decay (IBD) described by $\bar{\nu}_e + p \rightarrow e^+ + n$, with a mature technology of Gadolinium-loaded liquid scintillator \cite{Buck:2018cac}. The kinetic and annihilation energies deposited in the liquid by the positron form a prompt signal followed by the delayed signal of the neutron capture on gadolinium (17 $\mu$s mean capture time) and its subsequent de-excitation $\gamma$-cascade of 8 MeV total energy. 

With a mean reactor power of 52.76 $\pm$ 0.77 MW during the \textsc{Stereo} lifetime (Extended Data Fig.~\ref{fig:datataking}), the average rate of detected antineutrinos after all the selection cuts (see Methods section) is 394/day. For the \textsc{Stereo} experiment exposed to cosmic rays at the Earth surface and surrounded by other experiments using the ILL neutron beams, background rejection is a major challenge. Therefore, heavy lead and polyethylene shielding, distributed in the walls of the casemate and in the support structure around the detector, isolate the target volume from external $\gamma$ and neutron fluxes. Active rejection of the dominant cosmic ray background is also implemented: the top of the detector is covered by a muon veto and the pulse shape discrimination (PSD) capabilities of the liquid scintillator are exploited. The latter allows to reject the signals of fast neutrons coming from the interaction of cosmic rays around the detector. The combination of all these techniques brings the residual rate of accepted cosmic background events comparable to that of the antineutrinos. It is then precisely measured and subtracted thanks to the alternation of 50 days reactor cycles with nuclear fuel reloading or maintenance periods.

This ``ON-OFF'' subtraction implies a very precise control of the detector response throughout the whole data taking period. For this purpose, all the reactor and detector parameters are continuously monitored in order to correct for instabilities. Weekly calibration campaigns correct the slow drifts of the detector response and the same procedure of energy reconstruction is applied to the data and the simulation, which is a key tool to describe this response. In order to mitigate long-term effects the parameters of the simulation are fine-tuned to match calibration data at central dates of each of the two major data collection phases. Thus detection systematic uncertainties could be reduced down to the 1\% level (cf. Extended Data Fig.~\ref{fig:data-simul} and Methods section).

Here we present the \textsc{Stereo} results based on 107558 antineutrinos detected from October 2017 until the experiment was shut down in November 2020. This period gathers all the high quality data collected after the repair of detector defects in summer 2017 \cite{STEREO:2019ztb}. It corresponds to 273 reactor-ON days and 520 reactor OFF days, divided in two main data taking periods labelled as phase-II and phase-III. We discuss the impact of the complete analysis of our data (sterile neutrino rejection, neutrino deficit and spectral shape).

\section*{Rejection of the sterile neutrino hypothesis}\label{scn:oscillations}

The discrepancies between the predicted and observed antineutrino spectra render difficult the choice of an external oscillation-free reference spectrum for the sterile neutrino search. Thus, to realise a prediction-independent sterile neutrino search, in \textsc{Stereo} we leverage the significant dependence with the distance of the induced oscillation pattern for our baseline range (roughly 9 to 11 m). 
The different hypotheses $\mathcal{H}_\mu(\sin^2(2\theta_{ee}),\Delta m^2_{41})$ for the parameters of the sterile neutrino were tested by comparing the 6 independent antineutrino spectra recorded from the 6 cells to a common model where all experimental systematic uncertainties are propagated.
We adopt a profile $\Delta \chi^2$ approach and use as a test statistic
the difference between the minimum $\chi^2_{\sin^2(2\theta_{ee}),\Delta m^2_{41}}$ (where all but the $(\sin^2(2\theta_{ee}),\Delta m^2_{41})$ parameters are left free in the fit), and the global minimum $\chi^2_\mathrm{best~fit}$ (where the sterile neutrino parameters have also been optimised).
Since Wilk’s theorem conditions are not met~\cite{Neumair:2020elo}, we follow Feldman and Cousins’ prescription~\cite{Feldman:1997qc} and compute the $\Delta\chi^2$ distributions for each hypothesis using pseudo-experiments. When the p-value for the observed data given the hypothesis $\mathcal{H}_\mu(\sin^2(2\theta_{ee}),\Delta m^2_{41})$ is $< 0.05$, we exclude the corresponding $(\sin^2(2\theta_{ee}),\Delta m^2_{41})$ point in the
parameter space at the 95\% Confidence Level (CL). 

\begin{figure}
    \centering
    \includegraphics[width=\linewidth]{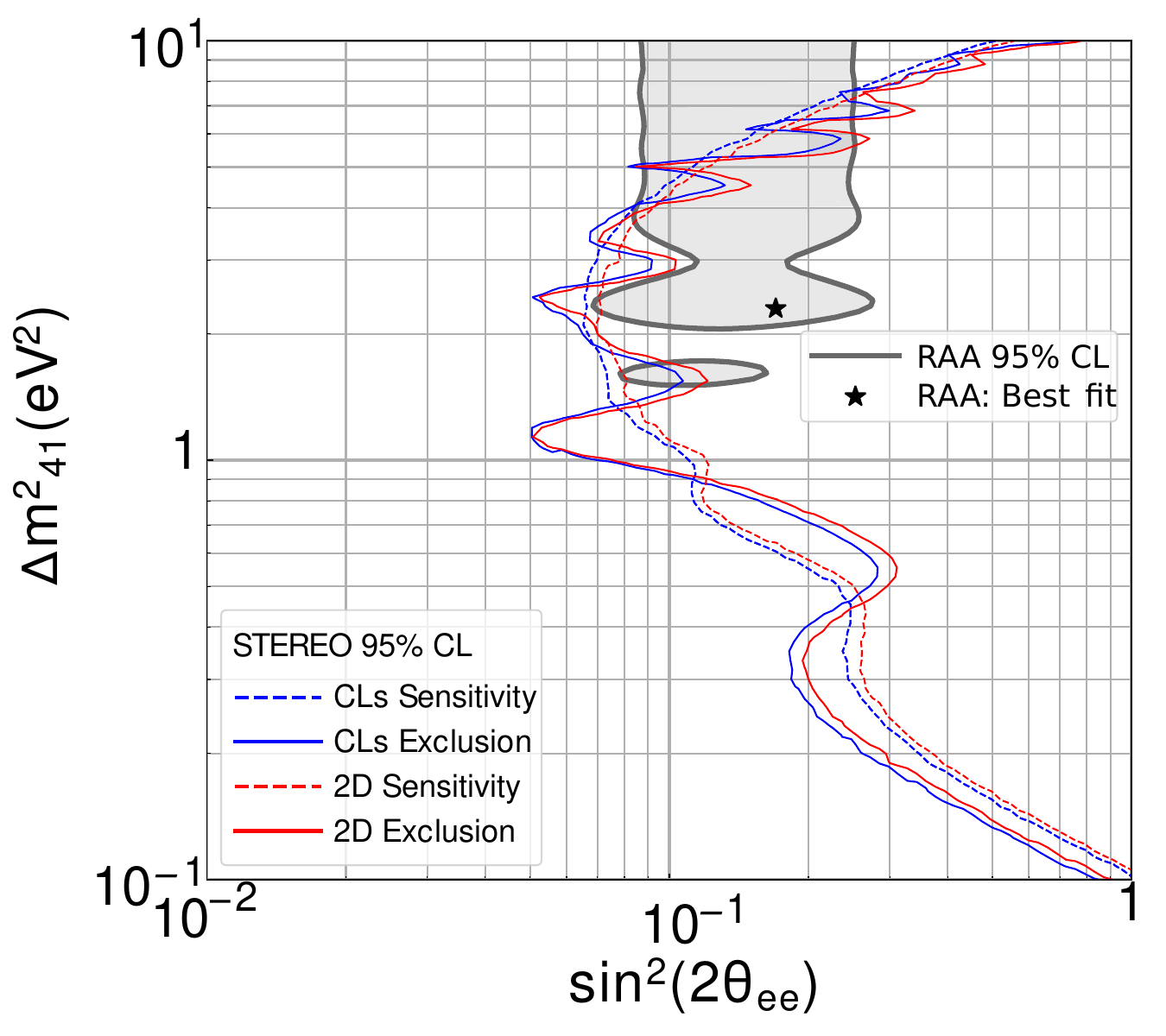}
    \caption{\textbf{Results of the \textsc{Stereo} oscillation analysis.} The exclusion contour (solid red) and exclusion sensitivity contour (dashed red) at 95\% CL are shown in the plane of the sterile neutrino parameters: the oscillation amplitude, $\sin^2(2\theta_{ee})$, related to sterile-active neutrinos mixing angle $\theta_{ee}$, and the oscillation frequency, $\Delta m^2_{41}$, which is also the difference of the square masses of the additional neutrino mass eigenstate and the lightest one. High values of $\sin^2(2\theta_{ee})$ are exluded (right of the exclusion curve). These results are derived in a 2D Feldman-Cousins framework (red). We verify our conclusions by following an alternative statistical prescription, the Gaussian CLs method~\cite{Qian:2014nha}, which yields a very similar exclusion contour (blue). The 95\% CL parameter space regions of the sterile neutrino as an explanation of the RAA (grey area) is rejected by \textsc{Stereo} below mass splittings of 4 eV$^2$. The initial RAA best fit point, marked by a star, is rejected with very high confidence level (p-value $< 10^{-4}$).
    }
    \label{fig:oscillation}
\end{figure}

Our analysis excludes the parameter space favored by the RAA up to a few eV$^2$ at 95\% CL or greater (see Fig.~\ref{fig:oscillation}). Additionally, we find that the best-fit points of the oscillation signals reported by the Neutrino-4~\cite{Neutrino4} and the Neos-Reno~\cite{RENO:2020hva} collaborations are excluded with significances of $3.3\sigma$ and $2.8\sigma$, respectively. We also observe that our data are compatible with the no oscillation hypothesis $\mathcal{H}(0,0)$ with a p-value of 0.52. Thus, the explanation of the Reactor Antineutrino Anomaly by a few eV mass sterile neutrino is strongly disfavored by STEREO data. Other experiments reach the same conclusion \cite{Andriamirado_2021,Alekseev:2022qxu}.

Interest in the high $\Delta m^2$ region (short wavelength oscillation) has been renewed recently by the results of the BEST experiment~\cite{Barinov:2021asz}, which confirm the Gallium anomaly. Sensitivity in this part of the parameter space is limited for \textsc{Stereo} due to the baseline ranges of our setup. Such sterile neutrino mass values are however in strong tension with Cosmic Microwave Background analyses~\cite{Planck:2018vyg}, and are probed by the KATRIN experiment~\cite{KATRIN:2022ith}.

\section*{Reference Uranium-235 antineutrino spectrum}\label{scn:spectrum}

\begin{figure*}[ht]
    \centering
    \includegraphics[width=\linewidth]{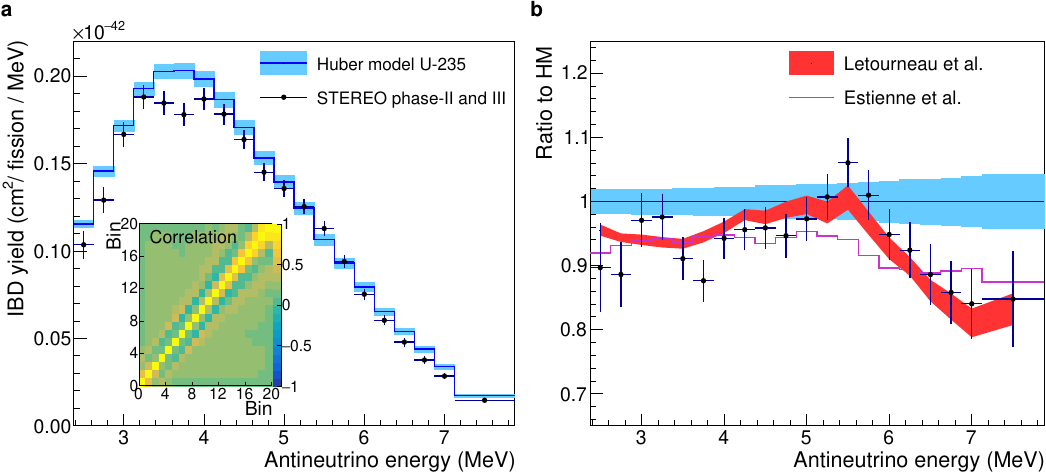}
    \caption{\textbf{New reference $^{235}$U antineutrino spectrum}. \textbf{a.} The unfolded antineutrino spectrum associated to the fission of $^{235}$U (black points) is shown with the HM prediction (blue) in the true antineutrino energy space. The vertical bars and blue band represent the respective total uncertainties at a 68\% CL (noted $1\sigma$ hereafter) and the vertical axis provides the absolute IBD yield. To obtain the HM prediction the emitted spectrum was multiplied by the theoretical IBD cross section \cite{STRUMIA200342}. The matrix illustrates the bin-to-bin correlations. Since the \textsc{Stereo} measurement is statistically limited, the pattern of correlations observed around the diagonal is mainly induced by the unfolding process. \textbf{b.} Relative deviations (black points) to the HM prediction (blue), exhibiting significant discrepancies in norm and in shape. However a better agreement is obtained with two recent summation models. The prediction of M. Estienne et al. \cite{Estienne:2019ujo} (magenta) corrects the evaluated nuclear data by including the most recent measurements of the $\beta$-strengths of the main fission products. It is in good agreement with the mean deficit measured by \textsc{Stereo} and could indicate the beginning of a shape distortion at high energy. A complementary approach \cite{Letourneau:2022kfs} (red band of $1\sigma$ uncertainties) generalises the correction of the $\beta$-spectra to all nuclei by completing the $\beta$-decay schemes of the ENSDF nuclear database \cite{ENSDF:2022} with a simple phenomenological Gamow-Teller $\beta$-decay strength model. Remarkable agreement with the STEREO spectrum is obtained both in normalisation and in shape.}
    \label{fig:spectrum}
\end{figure*}

The precise control of the detector response \cite{STEREO:2019ztb} combined with a good precision on the reactor power \cite{STEREO:2020fvd} allows to extend \textsc{Stereo}'s analysis to study the shape and absolute normalisation of the antineutrino spectrum. The highly enriched $^{235}$U fuel (HEU) used at ILL gives access to the antineutrino spectrum emitted by the fission of this single isotope, complementary to commercial reactors whose low enriched fuel (LEU) induces fission fractions distribution between $^{235}$U, $^{239}$Pu, $^{238}$U and $^{241}$Pu. As the 6 single-cell spectra are compatible with each other, the analysis of the antineutrino spectrum is performed by combining the selection of events in the 6 cells of the target volume. A spectrum is thus obtained as a function of the reconstructed energy deposited in the scintillator by the positron. However we provide a reference spectrum directly usable by the community by using a procedure to deconvolute the energy spectrum from the response of the detector. The method is based on a fit of the spectrum in antineutrino energy to the spectrum in reconstructed positron energy through the detector response matrix, the latter being derived from the Monte-Carlo simulation of the detector. The statistical fluctuations and the finite energy resolution of the detector induce numerical instabilities in this procedure. They are regularized by a Tikhonov type approach \cite{STEREO:2020hup,Stereo:2021wfd}. Biases of the method are encoded in a filter matrix provided in \cite{hepdata.132368}; applying this filter matrix to a spectrum prediction ensures an unbiased comparison with the unfolded \textsc{Stereo} spectrum, shown in Fig.~\ref{fig:spectrum}.a. This spectrum corresponds to the largest sample of pure $^{235}$U fission antineutrinos from a single experiment to date. The details of the systematic uncertainties relevant for this analysis are given in the Methods section.  

A specificity of \textsc{Stereo} is to control the absolute normalization of the measurement. Thus the integral of the spectrum provides the most precise measurement of the production of antineutrinos by the fission of $^{235}$U from a HEU reactor. It shows a deficit of $5.5\pm 2.1$ \% compared to the HM model, in excellent agreement with the world average (Extended Data Fig.~\ref{fig:lep_plot}), and confirms the prominent role of $^{235}$U in the RAA. 

Beyond this global deficit, the \textsc{Stereo} measurement also clearly highlights a spectral distortion compared to the HM model (Fig.~\ref{fig:spectrum}.b) that cannot be explained by an oscillation. There is thus a strong tension between this prediction, derived from the conversion of the total beta spectrum from $^{235}$U fissions, and that of the measured total antineutrino spectrum. A much better agreement with the \textsc{Stereo} results is obtained by the most recent summation models \cite{Estienne:2019ujo,Letourneau:2022kfs}. Such approach, complementary to the conversion (HM) method, builds a prediction with the sum of the antineutrino spectra of all fission products taken from the nuclear databases. A key ingredient of these models is the correction of the beta strengths involved in the decay of fission products. Experimentally it is difficult to measure beta transitions to the highly excited levels of the daughter nucleus, because of the high density of states and the complex gamma cascades that are associated \cite{HARDY1977307}. This results in a well-known bias in nuclear databases that tend to underestimate the contribution of low energy beta transitions. It seems that correcting this bias is a key element in bringing the summation models into agreement with the \textsc{Stereo} results \cite{Letourneau:2022kfs}.

\section*{Impact and Outlook}

The \textsc{Stereo} experiment provides the most accurate measurement to date of the antineutrino spectrum from pure $^{235}$U fission, corrected for the detector response. It is intended to be used by the particle physics community as a reference spectrum for future high precision reactor experiments, such as the determination of the mass hierarchy of neutrinos \cite{JUNO:2015zny} or the low-energy tests of the Standard Model with the since recently accessible process of Coherent Elastic Neutrino-Nucleus Scattering \cite{Abdullah:2022zue}. Observations of astrophysical \cite{Lagage1985} or geo-neutrinos \cite{Leyton2017} consider reactor antineutrinos as a background source and would also benefit from an improved description of their flux. For these purposes, all the results as well as the elements necessary to reproduce them are made publicly available \cite{hepdata.132368}.

We find significant deviation in normalisation and shape with respect to the HM prediction while at the same time we reject with high confidence level the hypothesis of a sterile neutrino of mass around 1 eV. Our findings support the conclusions of a recent global analysis of all neutrino data collected at reactors \cite{CONNIE:2022hna}, which points to a normalisation bias in the beta spectrum from $^{235}$U fission on which the HM prediction is anchored as the most likely explanation for the overall deficit. The involved nuclear data (neutron capture on lead and associated electron conversion) are being questioned. A recent direct measurement of the ratio of $^{235}$U to $^{239}$Pu fission beta spectra \cite{Kopeikin:2021ugh}, found to be 5\% lower than the ratio of the initial beta spectra \cite{Schreckenbach:1985ep,Hahn:1989zr}, supports this hypothesis. 

The comparison of our reference $^{235}$U spectrum with the most recent summation models illustrates a paradigm shift where a direct measurement of fission antineutrino spectra such as that of \textsc{Stereo} becomes a benchmark for nuclear data. Beyond its relevance for fundamental physics of neutrinos, this measurement of the total antineutrino spectrum from $^{235}$U has the potential to constrain the evaluated fission data with, in particular, a sensitivity to the fission yields \cite{Sonzogni:2016yac, Hayes:2015yka} and to the description of the beta transitions of fission products \cite{Estienne:2019ujo, Letourneau:2022kfs}. Our reference spectrum as well as the associated gain in reliability and accuracy of the nuclear data finds direct applications in the operation and surveillance of reactors \cite{Bernstein:2019hix}.

\section*{The \textsc{Stereo} collaboration}

\noindent
H.~Almaz\'an$^{1,6}$,
L.~Bernard$^{2}$,
A.~Blanchet$^{3,7}$,
A.~Bonhomme$^{1,3}$,
C.~Buck$^{1}$,
A.~Chalil$^{3}$,
P.~del~Amo~Sanchez$^{4}$,
I.~El~Atmani$^{3,8}$,
L.~Labit$^{4,9}$,
J.~Lamblin$^{2}$,
A.~Letourneau$^{3}$,
D.~Lhuillier$^{3,*}$,
M.~Licciardi$^{2}$,
M.~Lindner$^{1}$,
T.~Materna$^{3}$,
H.~Pessard$^{4}$,
J.-S.~R\'eal$^{2}$,
J.-S.~Ricol$^{2}$,
C.~Roca$^{1}$,
R.~Rogly$^{3}$,
T.~Salagnac$^{2}$,
V.~Savu$^{3}$,
S.~Schoppmann$^{1,10}$,
T.~Soldner$^{5}$,
A.~Stutz$^{2}$,
and M.~Vialat$^{5}$

~

\noindent
$^{1}$~Max-Planck-Institut f\"ur Kernphysik, Saupfercheckweg 1, 69117 Heidelberg, Germany \\
$^{2}$~Univ. Grenoble Alpes, CNRS, Grenoble INP, LPSC-IN2P3, 38000 Grenoble, France \\
$^{3}$~IRFU, CEA, Universit\'e Paris-Saclay, 91191 Gif-sur-Yvette, France \\
$^{4}$~Univ. Savoie Mont Blanc, CNRS, LAPP-IN2P3, 74000 Annecy, France \\
$^{5}$~Institut Laue-Langevin, CS 20156, 38042 Grenoble Cedex 9, France \\
$^{6}$~Present address: University of Manchester. Schuster Building. M139PL, United Kingdom\\
$^{7}$~Present address: LPNHE, Sorbonne Universit\'e, Universit\'e de Paris, CNRS/IN2P3, 75005 Paris, France\\
$^{8}$~Present address: Hassan II University, Faculty of Sciences, A\"in Chock, BP 5366 Maarif, Casablanca 20100, Morocco\\
$^{9}$~Present address: Univ. Bordeaux, CNRS, LP2i Bordeaux, UMR 5797, F-33170 Gradignan, France\\
$^{10}$~Present address: University of California, Department of Physics, Berkeley, CA 94720-7300, USA and Lawrence Berkeley National Laboratory, Berkeley, CA 94720-8153, USA\\  

\noindent
$^{*}$~Corresponding author: david.lhuillier@cea.fr\\

\clearpage


\clearpage
\section*{Methods}

\subsection*{Data taking}

The \textsc{Stereo} experiment has taken data from November 2016 to November 2020. In the first phase of the experiment (Nov. 2016 - Feb. 2017) the response of the detector was affected by several defects of the acrylic components of the inner detector: the light collection in cell 4 and in the front gamma-catcher was reduced by a factor about 2.5 due to a leak of mineral oil optically coupling the PMTs and the acrylic buffers; the cross-talk between cells significantly increased over time due to ingress of liquid scintillator inside the sandwich structure of most acrylic walls. Taking advantage of a long maintenance period of the reactor the detector was repaired during summer 2017 and the subsequent phase-II (Oct. 2017 - Apr. 2019) and phase-III (Apr. 2019 - Nov. 2020) of data taking were shown to be far more stable with very similar responses from all target cells. All results reported here are based on the high quality data from these last two phases (see Extended Data Fig.~\ref{fig:datataking}) \cite{stereo:2018:1,stereo:2018:2,stereo:2018:3,stereo:2018:4,stereo:2019:0,stereo:2019:1,stereo:2019:2,stereo:2020:1,stereo:2020:2}. They contain about twice as much data as in previous \textsc{Stereo} publications \cite{STEREO:2019ztb,STEREO:2020fvd,STEREO:2020hup,STEREOOscillationPhase1only}.

Weekly calibration runs are performed by inserting a $^{54}$Mn source at 5 different heights above the bottom of the target ($z=$10~cm, 30~cm, 45~cm, 60~cm and 80~cm) in each of the 6 cells, with the exception of cell 3 where a similar tube is present but used to fill the liquid scintillator. In addition, full calibration campaigns are performed every few months with a complete set of sources ($^{137}$Cs, $^{54}$Mn, $^{65}$Zn, $^{42}$K, $^{60}$Co, $^{24}$Na and $^{241}$Am-$^{9}$Be) covering the neutrino energy range and inducing various topologies of energy deposits (single $\gamma$, multi-$\gamma$, $\gamma$+neutron). Some of these sources had much higher rate than antineutrino interactions. In order to be sure that the baseline voltage of the electronics has been recovered between every two events, only events with at least $10~\mu$s to the previous event are considered in the analysis of calibration runs.

\subsection*{Detector response}

The precise control of the STEREO detector response is a prerequisite to extract all observables for neutrino physics. The accurate description of the detector response in the GEANT4 \cite{G4} simulation, which describes propagation and interaction of particles in the detector, the scintillation process including quenching, the light transport and conversion into photoelectrons in the PMTs \cite{STEREO:2019ztb}, is thus validated with a dataset as complete as possible and independent of the neutrino data. In addition to calibration data from sources, two cosmic rays interactions were also studied: the $\gamma$-peak at 2.2 MeV induced by the capture of spallation neutrons, created predominantly in the surrounding lead shielding, on hydrogen; and the $\beta$-decay spectrum of $^{12}$B nuclei, generated by muon capture on $^{12}$C in the scintillator. These cosmogenic events are more uniformly distributed in the detector volume than source events (produced around calibration tubes) and provide complementary information. The calculation of the beta spectrum of $^{12}$B has been updated with respect to previous analyses \cite{STEREO:2020hup} by including more complete calculations of the correction terms to Fermi theory. The prediction of the BetaShape code \cite{Mougeot2017} is chosen as the new reference and provided in the supplementary materials \cite{hepdata.132368}.

\subsubsection*{Energy scale}
The quality of the energy scale is estimated from the residuals between data and simulation using all the control data mentioned above. For radioactive sources it is the position of the reconstructed $\gamma$-peaks that is used, whereas in the case of $^{12}$B it is the shape of the beta spectrum. These two pieces of information are combined in a global fit of the possible relative distortions between the energy scales of the data and the simulation. The source residuals provide a direct measure of these distortions. Their propagation in the shape of the beta spectrum, more complex, is described by the formalism presented in \cite{Mention:2017dyq}. Extended Data Fig.~\ref{fig:data-simul} shows all residuals at the target level (averaged over all cells) with an example of global fit using a polynomial of order three as a model for the distortions. Higher order polynomial as well as a very general Kernel Density Estimation (KDE) approach are also used. The energy scale uncertainty for any given cell is found to be 1\%, not correlated between cells, and 0.4\% at the target level. The same results apply to both data collection phases.\\

\subsubsection*{Stability of energy reconstruction}
As the detector response evolves with time, two fine-tunings of the simulation are performed to align the simulated to the experimental distributions of charges on central dates of each phase (28/04/2018 for phase-II and 28/02/2020 for phase-III). The evolution of the detector before, between and after these tuning dates is accurately monitored with the weekly $^{54}$Mn calibration runs by computing calibration coefficients (CCs) using the formalism described in \cite{STEREO:2019ztb}. Extended Data Fig.~\ref{fig:nH}.a shows a continuous decrease of collected photo-electrons in each cell for the same energy deposits of the source. The 10 to 20\% amplitude across the whole duration of the experiment is mainly attributed to the loss of transparency of the liquid scintillator. The monitoring of the position of the n-H peak from cosmic events, performed for the same period of time using this set of CCs  (Extended Data Fig.~\ref{fig:nH}.b), demonstrates a remarkable time stability of the energy reconstruction with 0.25\% associated uncertainty ($1\sigma$ level) achieved for each cell. The decrease of collected photo-electrons results in a slow degradation of the energy resolution (Extended Data Fig.~\ref{fig:nH}.c).

\subsubsection*{Stability of Pulse Shape Discrimination variable}

The Pulse Shape Discrimination (PSD) variable is an important observable used to discriminate electron-like and proton-like energy deposits, the prompt signal of an IBD event being electron-like while the dominant fast neutron background induced by cosmic rays being proton-like. The liquid scintillator used in \textsc{Stereo} has the ability to perform Pulse Shape Discrimination: the time distribution of emitted scintillation photons, and of the subsequent charge output of the PMTs, depends on the density of energy deposition in the liquid. Proton recoils produce excited states of the scintillator molecules that have longer de-excitation time, leading to an enhancement of the late part ($Q_\mathrm{tail}$) of the output electronic pulse (cf. Extended Data Fig.~\ref{fig:PSD}.a). The PSD variable, defined as the fraction of charge in the late part of the pulse: $Q_\mathrm{tail}/Q_\mathrm{tot}$, therefore discriminates between electron-like deposits and proton-like deposits \cite{Buck:2018cac}. The discrimination between these two populations can be seen with antineutrino data on Extended Data Fig.~\ref{fig:SB}.b.

However the shape of the light pulses is sensitive to the time evolution of the liquid scintillator properties. Increase in temperature dilates the liquid and reduces the effective density of energy deposition shifting the PSD variable to lower value. Degradation of the optical properties of the liquid (transparency) reduces the separation of the e-like and p-like populations. An example of the discrepancy of PSD distributions between the beginning and end of phase-III is displayed on Extended Data  Fig.~\ref{fig:PSD}.b.

The evolution of the PSD variable is not taken into account by the calibration strategy presented in the previous paragraph. It requires a dedicated correction, based on the Am-Be source (n+$\gamma$ type) that has a PSD spectrum comparable to that of antineutrinos. The principle is the following: Am-Be calibration data, taken every 1-2 months, are used to monitor the PSD variable (in each cell independently); the parameters ruling the PSD evolution with time and temperature are fitted to Am-Be data to provide a continuous correction of the PSD variable; this correction is applied to antineutrino runs before the extraction of the IBD signal. 

The distortion of the PSD at date $t$ with respect to a reference date $t_\mathrm{ref}$ (taken in the middle of the data taking: May 28th, 2019) is expressed with a scaling factor $f$ and a shift $S$: \begin{equation}
    \mathrm{PSD}(t) = f \times \mathrm{PSD}(t_\mathrm{ref}) + S.
\end{equation}
The quantities $f$ and $S$ are determined for each Am-Be calibration run, every 1-2 months. The continuous evolution is obtained by a fit with a polynomial function of time (second-order) and temperature (first order):\begin{equation}
  \left\{ \begin{array}{c}
     f(t,T) = a_f t + b_f t^2 + c_f T + d_f \\
       S(t,T) = a_s t + b_s t^2 + c_s T + d_s
  \end{array}  \right. .
\end{equation} 
The values of $S$ from Am-Be runs and the fitted $S(t,T)$ are shown for illustration in Extended Data Fig.~\ref{fig:PSD}.c for cell 4. The expression of the fit function was chosen to be as simple as possible. A first-order polynomial of the temperature provides a good fit to the data (Extended Data Fig.~\ref{fig:PSD}.d1); however, a linear dependence on time was not sufficient to correctly describe the evolution of the PSD variable (Extended Data Fig.~\ref{fig:PSD}.d2).

It was found that the fit coefficients $(a_f,b_f,c_f,d_f)$ and $(a_s,b_s,c_s,d_s)$ do not significantly depend on either the height of the source in the calibration tube or the energy of the events. Therefore the correction was build with calibration runs in the center of each cell ($z=45$~cm), combining events from a wide energy range: from 1.625 to 4.125 MeV.
The value of the PSD variable for each physics run (either reactor-on or -off) at date $t$, with detector temperature $T$, is then corrected based on the fitted $f(t,T)$ and $S(t,T)$. 

\subsection*{Signal and backgrounds}

The extraction of the inverse $\beta$-decay (IBD) signal relies on two steps. First, the cuts listed in Extended Data Fig.~\ref{fig:SB}.a are applied: they use the topology of IBD events, with a prompt signal ($e^+$ kinetic energy deposit and annihilation) followed by a delayed signal ($n$ capture). There are four types of cuts: i) cuts defining the energy range of the prompt and delayed events, ii) cuts to select pairs of prompt-delayed events with appropriate time and space coincidence, iii) topology cuts to ensure that most of the prompt energy is deposited in a single target cell (cells 1-6), and iv) cuts removing cosmogenic background. The cut on PMT asymmetry, for instance, requires the charge to be evenly distributed among all 4 PMTs of the vertex cell, thus removing events happening very close to a single PMT such as muon decays. The selection cuts are applied on both reactor-on (``ON'') and reactor-off (``OFF'') samples. The amount of random coincidences passing these cuts is estimated using an off-time method. A more complete description of all cuts can be found in \cite{STEREO:2019ztb}.

In the second step, a joint fit of ON and OFF PSD distributions is performed in each energy bin; the difference between ON and OFF, fitted by a Gaussian, gives the number of IBD candidates (Extended Data Fig.~\ref{fig:SB}.b) \cite{STEREO:2019ztb}.
Event rates are extracted separately for each phase of data taking in order to suppress residual effects from detector ageing. The average signal-to-background ratio is 0.8 for phase-II and 1.0 for phase-III; the higher value for phase-III is due to higher mean reactor power. Extended Data Fig.~\ref{fig:SB}.c displays the energy-dependent signal-to-background ratio: it is limited by random coincidences at low energy and by the decrease in signal intensity at high energy.
For the oscillation analysis, the procedure described above provides 12 spectra (6 cells $\times$ 2 phases) with 11 energy bins (500~keV wide, from 1.625~MeV to 7.125~MeV). Data from the 6 target cells are merged for the spectrum analysis giving 2 input spectra (one per phase) with finer energy binning (22 bins of 250~keV).

The PSD fit method only accounts for cosmogenic backgrounds that are monitored when the reactor is turned off. To check for correlated background related to reactor operation we directly compared the ON and OFF PSD distributions corrected for the atmospheric pressure instead of fitting a relative normalisation parameter. Any discrepancy could be an indication of backgrounds from the reactor or other instruments neighboring \textsc{Stereo} in ILL's reactor hall. A hint of an excess of events is found at low energy for the two phases, as shown in Extended Data Fig.~\ref{fig:SB}.d. It is fitted by a power law, which is used to correct the neutrino rates (few \% correction at low energy, negligible above 3 MeV). However, the origin of this reactogenic background is unclear. It may be produced by fast neutrons, inducing proton recoils in the detector. Such events would not be confused with the $e^+$ annihilation from the IBD signal thanks to Pulse Shape Discrimination; therefore, a 100\% uncertainty is added to this background component.

For a global validation of the extraction of neutrino rates the \textsc{Stereo} dataset has been divided into seven ``mini-experiments'' each consisting of 1 reactor cycle flanked by OFF reactor periods. Extended Data Fig.~\ref{fig:ONcomp} shows that the 7 neutrino spectra thus obtained are compatible with each other with a distribution of fluctuations as expected from pure statistics. This cross-check rules out significant spectrum distortions that could be induced by residual effects of the time evolution of the detector response or background.

\subsection*{Neutron efficiency}

An accurate computation of the selection efficiencies is a key aspect to normalize the experimental rates and compare them to flux predictions. These efficiencies are computed with the Monte-Carlo simulation of the experiment. However, the description of low-energy neutron physics may be imperfect in this simulation; therefore, the efficiency associated to tagging the delayed signal (called \textit{neutron efficiency}) is validated with calibration data. The $^{241}$Am-$^{9}$Be source, producing correlated pairs ($\gamma$,n) mimicking IBD events, is used to evaluate this efficiency. Its deployment into the target cells at five different heights allows to probe the entire detector volume.

All selection cuts of the delayed signal described in Table 1 are applied, except the 600~mm distance between the prompt and delayed vertices, not relevant for this study. The resulting amount of ($\gamma$,n) pairs is compared to a sample without these cuts, giving the neutron efficiency for measured and simulated events.

A precise modeling of the de-excitation of Gd isotopes after neutron capture is obtained using the FIFRELIN code \cite{FIFRELIN}. New features have been implemented to improve the reliability of the \textsc{Stereo} detector response compared to the previous version \cite{STEREO_FIFRELIN}: the physics of conversion electrons and the subsequent X-ray emission has been treated more accurately; a better description of the high-energy part of the $\gamma$-ray spectrum has been achieved by taking into account available experimental data on primary $\gamma$-rays; a complete treatment of angular correlations between $\gamma$-rays has been implemented \cite{fifrelin_2022}.

Due to the high intensity of our Am-Be source inducing a 27~kHz event rate in the detector, many random coincidences have to be removed to get a clean sample of ($\gamma$,n) pairs. They are estimated with an off-time windows search and subtracted statistically. The estimation was improved with respect to the work presented in \cite{STEREO:2019ztb}. The procedure has been validated with two Am-Be simulations: one with high intensity (27~kHz, similar to the data) and another at low intensity (1~Hz) where random coincidences are negligible. 

The neutron efficiency is displayed as function of the source position in $(X, Z)$ in Extended Data Fig.~\ref{fig:neuteff}.
The spatial dependence of the efficiency is well reproduced in the simulation, although there is a $\sim$1.5\% global difference between data and simulation. A 3D spatial model of the efficiency $\varepsilon_\mathrm{n}(X,Y,Z)$ is built by fitting the Am-Be points in data or simulation, assuming the same behaviour in the two horizontal dimensions ($X$ and $Y$). The coefficient \begin{equation}c_\mathrm{n} (X,Y,Z) = \varepsilon_\mathrm{n}^\mathrm{data}(X,Y,Z)/\varepsilon_\mathrm{n}^\mathrm{MC}(X,Y,Z).\end{equation}
corrects the bias on the selection efficiency due to the imperfect modelling of neutron physics in simulation.
The integration of $c_\mathrm{n}$ over each cell volume gives a correction factor for the amount of IBD candidates detected in each cell. The cell-averaged coefficient $\overline{c_\mathrm{n}}$ is the same for all cells: \begin{equation}
\begin{tabular}{c@{ }l}
   \multirow{2}{*}{$\overline{c_\mathrm{n}} = $} & $0.9824 \pm 0.0063$ [uncorr] $\pm 0.0019$ [corr] phase-II,\\
 & $0.9847 \pm 0.0063$ [uncorr] $\pm 0.0027$ [corr] phase-III,\\
    \end{tabular}
\label{eq:cneut}
\end{equation}
where part of the uncertainty is correlated [corr] between cells (change in $\varepsilon_\mathrm{n}^\mathrm{MC}$ due to source intensity in simulation, discrepancies on $\overline{c_\mathrm{n}}$ due to the choice of the 3D spatial model) and part is uncorrelated [uncorr] between cells (statistical error on $\varepsilon_\mathrm{n}$ measurements, uncertainty on the positions of the Am-Be source in each calibration tube).

\subsection*{Sterile neutrino search}

The data recorded in \textsc{Stereo}’s 6 cells are binned in 11 equally wide 500 keV energy bins from 1.625 to 7.125 MeV in reconstructed energy, separately for the running periods phase-II and III, and are fitted to the predicted spectra by minimising the following $\chi^2$: 
\begin{equation}\begin{aligned}
&\chi^2\left(\sin^2(2\theta_{ee}), \Delta m^2_{41}; \phi_i, \alpha^j \right)= \\
\sum_{p=\text{II}}^{\text{III}} \sum_{l=1}^{N_{\text{cells}}} 
\sum_{i=1}^{N_{\text{Ebins}}} &\left( \frac{D_{p,l,i}-\phi_i\, M_{p,l,i}(\sin^2(2\theta_{ee}),\Delta m^2_{41}; \alpha^j) }{\sigma_{p,l,i}}\right)^2 \\
+\sum_{l=1}^{N_{\text{cells}}}& \left( \frac{\alpha_l^{\text{EscaleU}}}{\sigma_l^{\text{EscaleU}}} \right)^2 
+ \left( \frac{\alpha_l^{\text{NormU}}}{\sigma_l^{\text{NormU}}} \right)^2 
+ \left( \frac{\alpha_l^{\text{ReactBg}}}{\sigma_l^{\text{ReactBg}}} \right)^2\\
+&\left( \frac{\alpha^{\text{EscaleC}}}{\sigma^{\text{EscaleC}}} \right)^2
+\left( \frac{\alpha^{\text{Cuts}}}{\sigma^{\text{Cuts}}} \right)^2
+\left( \frac{\alpha^{\text{II VS III norm}}}{\sigma^{\text{II VS III norm}}} \right)^2 
\label{eqn:chi2}
\end{aligned}\end{equation}
where  $D_{p,l,i}$ stands for the measured IBD rate in the $i$-th energy bin of cell number $l$ for running phase $p$, $M_{p,l,i}(\sin^2(2\theta_{ee}), \Delta m^2_{41}, \alpha^j)$ is the predicted IBD rate in the corresponding bin for the sterile neutrino parameters ($\sin^2(2\theta_{ee}),\Delta m^2_{41}$), and $\alpha^j$ are the nuisance parameters. The $\phi_i$ variables apply a cell-independent correction to the predicted spectra of each cell $M_{p,l,i}$, thus absorbing any possible difference between the spectrum prediction and the observed reactor antineutrino spectrum, rendering the analysis prediction-independent. 
The statistical uncertainty $\sigma_{p,l,i}$, stemming from the background subtraction and the signal statistics, is parameterised as a function of the expected number of signal events, which depends on the values of $\phi_i$ and the sterile neutrino parameters. Indeed, some values of these parameters lead to predicted rates $M_{p,l,i}$ that are significantly different than the measured values, preventing to use the statistical uncertainty directly obtained from the signal extraction.  
Finally, the nuisance parameters $\alpha^j$ address the systematic uncertainties by allowing for adjustments of the predicted rates within the constraints given by auxiliary measurements of detector response and backgrounds. These constraints are at the percent or sub-percent level and are either uncorrelated (part of energy scale $\alpha_l^{\text{EscaleU}}$, normalisation $\alpha_l^{\text{NormU}}$, reactor background $\alpha_l^{\text{ReactorBg}}$) or correlated (part of energy scale $\alpha^{\text{EscaleC}}$, selection cuts $\alpha^{\text{Cuts}}$, relative normalisation of phases $\alpha^{\text{II~vs~III~norm}}$) between cells. Good systematics control is paramount in the low $\Delta m^2$ region as an energy scale shift can mimic slow oscillations. A summary of uncertainties can be found in Extended Data Table~\ref{tab:syst}.a.

The $\Delta \chi^2$ distributions required for the statistical analysis of a given hypothesis ($\sin^2(2\theta_{ee})$, $\Delta m^2_{41}$) are computed by generating pseudo-experiments and fitting them with the $\chi^2$ of Eq.~(\ref{eqn:chi2}) (Feldman-Cousins approach). The pseudo-experiments are produced by taking the predicted spectrum in each cell and independently fluctuating each energy bin according to a normal distribution whose standard deviation is the statistical uncertainty of the bin $\sigma_{p,l,i}$. Variations within the systematic uncertainties are also included.
A detailed account of the statistical analysis leading to the exclusion contour in the sterile neutrino parameter space ($\sin^2(2\theta_{ee})$, $\Delta m^2_{41}$) is given in~\cite{STEREO:2019ztb}. 

Simultaneous optimisation of $\phi_i$ and $\alpha^j$ while fixing $\sin^2(2\theta_{ee})$ and $\Delta m^2_{41}$ to zero allow us to compare our data to the predicted spectrum for the no sterile neutrino hypothesis. This hypothesis is a good fit of the data (goodness-of-fit is 25\%), as can be seen in Extended Data Fig.~\ref{fig:cell_spectra}.

A further independent sterile neutrino search was performed as a crosscheck~\cite{labit:tel-03596718}. An analytical parameterization of the detector response was built for the phase-II period. The systematic uncertainties on this detector response model were determined by fitting the cosmogenic $^{12}$B data and comparing the resulting response model to the one obtained from the Geant4 simulation. The phase-II data, binned in 250 keV bins, was then fitted to the predicted spectra for each cell. The prediction was computed on the fly by the software by convolving a common reactor antineutrino spectrum with each cell's detector response function. This approach allows for a more natural treatment of the systematics, detector- as well as spectrum-related, since the corresponding parameters can be left free to vary during the fit. Thus, all reactor spectrum bins between 2.75 and 7.25 MeV in antineutrino energy were free to vary in the fit. Additionally, 15 detector response parameters per cell were left free in the fit, including 3 for energy-scale and 4 concerning the energy resolution. The resulting exclusion contour was found to be compatible with the published analysis of the phase-II data~\cite{STEREO:2019ztb}. 
  
\subsection*{Spectrum analysis}
The systematic uncertainties relevant for the spectrum analysis are given in Extended Data Table.~\ref{tab:syst}.b. A survey of the diaphragm used to determine the flow of heavy water inside the primary circuit of the ILL reactor has been performed in 2022. It has shown an excellent stability of the geometry of the aperture, thus confirming the accuracy of the evaluation of the reactor power (1.4\% uncertainty \cite{STEREO:2020fvd}).

Two complementary implementations of the Tikhonov approach developped by the \textsc{Stereo} collaboration give the same results. They are presented in details in \cite{STEREO:2020hup,Stereo:2021wfd}. The strength of the Tikhonov regularisation is carefully chosen to preserve the shape information of the spectrum at the scale of a few bins, but by definition it dampens the fluctuations between neighbouring bins, which become correlated. For the same reason, a real high-frequency (2-bin) feature in the spectrum will be smoothed out. These two effects are encoded in the covariance matrix and in a filter matrix, available in the supplementary materials \cite{hepdata.132368}. Importantly, models should be folded through the filter matrix to ensure an unbiased comparison with \textsc{Stereo} unfolded spectrum \cite{Stereo:2021wfd}.

The HM-based prediction for the IBD rates is built as follows. The antineutrino flux ($\overline{\nu}$/fission) is taken as the exponential of a sixth-order polynomial based on Ref.~\cite{Huber:2011wv}. We then multiply the antineutrino flux by the IBD cross section from Ref.~\cite{STRUMIA200342}. 
In order to characterize the spectral distortion around 5.5~MeV, we describe the excess by a Gaussian and build the following model based on the HM prediction $\Phi^\mathrm{HM}$: 
\begin{equation}\label{eqn:bump-model}
    M(E_\nu) = a \cdot \Phi^\mathrm{HM}(E_\nu) \left[ 1+ A \exp -\frac{(E_\nu - \mu)^2}{2\sigma^2} \right]
\end{equation}
The comparison with \textsc{Stereo} spectrum is done by passing the prediction (\ref{eqn:bump-model}) through the filter matrix. The fit with the four parameters improves the $\chi^{2}$ by 29.0 units compared to the fit with only the normalization offset, corresponding to a statistical significance of 4.6$\sigma$ for the 5.5~MeV distortion. The maximum amplitude of the Gaussian $A$ is fitted to be (15.6 $\pm$ 5.2)\%.
This distortion was initially observed by experiments with power reactors with similar parameters. Our result for pure $^{235}$U thus indicates that a similar bump is probably present in the fission spectrum of $^{239}$Pu. Based on the summation model \cite{Letourneau:2022kfs} the distortion in the neutrino spectrum would imply a similar distortion in the beta spectrum. As a possible explanation of why it was not seen in the reference fission beta spectrum measured in the 1980s we suggest that it could be due to a slight change in the calibration of the magnetic spectrometer \cite{MAMPE1978127} during the scan of the 2-9 MeV energy range. The high sensitivity to such a kink in energy scale has been discussed in \cite{Mention:2017dyq}.

\backmatter
\section*{Data availability}

All the results (sterile neutrino search and spectrum analysis) and the elements necessary to reproduce them are provided in the supplementary materials available at Ref.~\cite{hepdata.132368}. In accordance with the ILL data policy, all raw data are available at \cite{stereo:2018:1,stereo:2018:2,stereo:2018:3,stereo:2018:4,stereo:2019:0,stereo:2019:1,stereo:2019:2,stereo:2020:1,stereo:2020:2}. Additional information is available upon request by contacting the \textsc{Stereo} Collaboration.

\bmhead*{Acknowledgements}

This work is funded by the French National Research Agency (ANR) within the project ANR-13-BS05-0007 and the ``Investments for the future'' programme ENIGMASS LabEx (ANR-11-LABX-0012). Authors are grateful for the technical and administrative support of the ILL for the installation and operation of the \textsc{Stereo} detector. We further acknowledge the support of the CEA, the CNRS/IN2P3 and the Max Planck Society.\\

\section*{Declarations}

\bmhead*{Contributions}

All listed authors have contributed to the present publication. The different contributions span from the design and construction of the \textsc{Stereo} detector and its installation on the reactor site, to the acquisition of data and the development of the simulation and analysis software. The manuscript was reviewed by the whole collaboration$^{\dagger}$ and all authors approved its final version; the authors’ names are listed alphabetically.

Corresponding author: David Lhuillier.

$^{\dagger}$~\url{https://www.stereo-experiment.org}\\

\bmhead*{Competing interests}

The authors declare no competing interests.

\begin{appendices}

\clearpage

\section*{Extended Data}

\makeatletter
\@ifundefined{@figurecaption}%
{\renewcommand{\floatc@plain}[2]{{\bfseries#1}\hskip.7em#2\par}}%
{\renewcommand{\floatc@plain}{\@figurecaption}}
\makeatother
\floatstyle{plain}
\newfloat{extdatafigure}{hp}{exf}
\floatname{extdatafigure}{Extended Data Fig.}
\newfloat{extdatatable}{hp}{ext}
\floatname{extdatatable}{Extended Data Table}

\begin{extdatafigure*}
    \centering
    \includegraphics[width=0.9\linewidth]{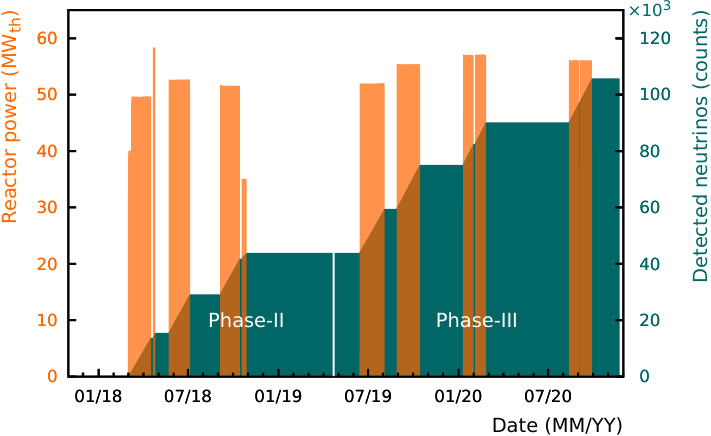}
    \caption{\textbf{\textsc{Stereo} data taking}. The left-hand axis refers to the reactor power graph (orange) while the associated cumulative number of detected antineutrinos (dark green) can be read on the right-hand axis. Three reactor cycles occurred during phase-II and four other cycles in phase-III. The alternation of reactor-on (``ON'') and reactor-off (``OFF'') periods is a key aspect of the experiment in order to accurately control the subtraction of cosmogenic background.}
    \label{fig:datataking}
\end{extdatafigure*}

\begin{extdatafigure*}
    \centering
    \includegraphics[width=\linewidth]{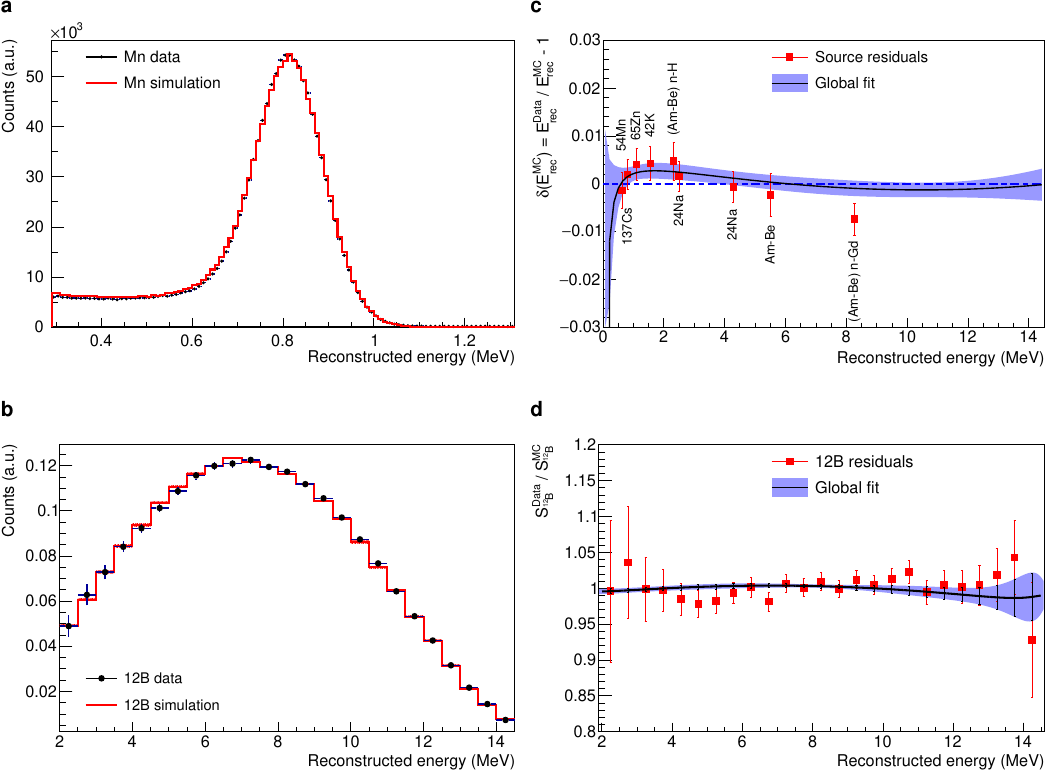}
    \caption{\textbf{Accurate control of the energy scale.} \textbf{a.} Experimental and simulated reconstructed energy spectra of the $^{54}$Mn source, with the source at 45 cm above the bottom of cell 4. \textbf{b.} Experimental and simulated $^{12}$B beta spectra. Only statistical uncertainties are displayed in these two first plots. \textbf{c.} Relative difference of the experimental and simulated position of $\gamma$-peaks from various radioactive sources (red points). For multi-$\gamma$ sources, all photons are reconstructed in the same event. n-H and n-Gd peaks originate from the Am-Be source. The reconstructed energies differ from the physical ones because of the quenching effect. The statistical uncertainties are negligible for both the simulation and the data, the dominant contribution comes from the systematics in the determination of the peak positions (including time stability, choice of fit function and fit range). \textbf{d.} Ratio of measured and simulated $^{12}$B spectrum (red points). The error bars include the systematic uncertainties of the simulated $^{12}$B spectrum (dominant contribution) and the statistical uncertainty of the measured spectrum. In \textbf{c} and \textbf{d} the black line and blue band correspond to the global fit of all residuals and its associated uncertainty, respectively. Here a third order polynomial is used as a model of the relative distortions between the experimental and simulated energy scales. This function applies directly to the residuals in \textbf{c} while it is converted into spectrum shape distortions to fit the points in \textbf{d} following the formalism described in \cite{Mention:2017dyq}. All uncertainties are taken at a 68\% CL (noted $1\sigma$ hereafter). Only phase-III results are displayed here, similar results are obtained for phase-II \cite{STEREO:2019ztb}.}
    \label{fig:data-simul}
\end{extdatafigure*}

\begin{extdatafigure*}
    \centering
    \includegraphics[width=0.60\linewidth]{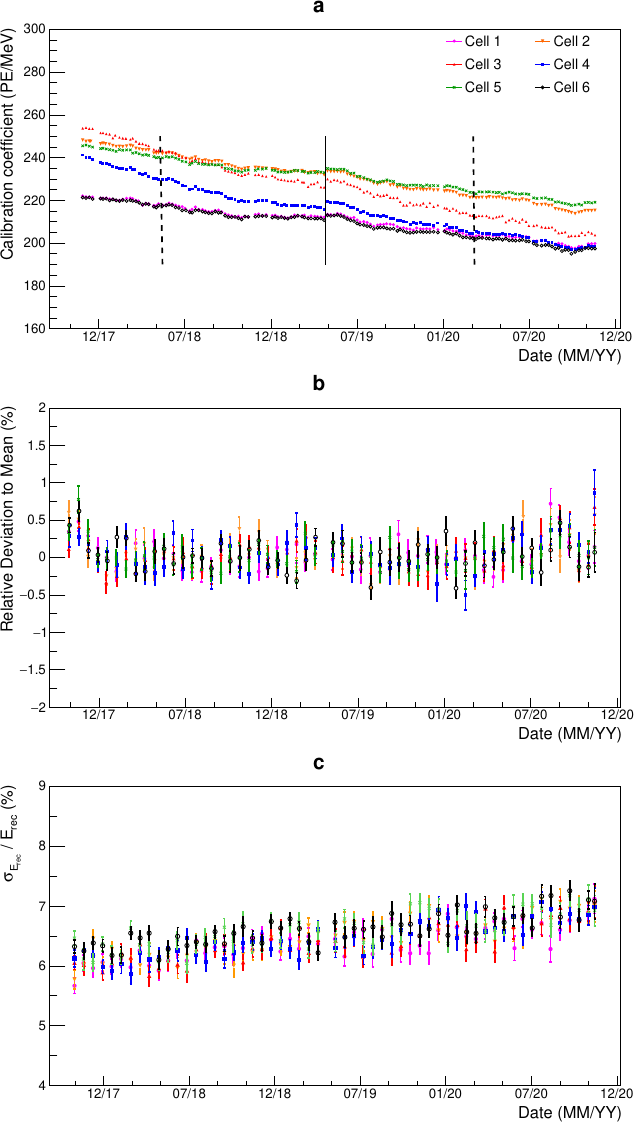}
    \caption{\textbf{Time evolution of the detector response.} \textbf{(a.)} Evolution of the calibration coefficients (CCs) for the 6 target cells of the \textsc{Stereo} detector. The decrease over time mostly can be explained by a reduction of the attenuation length of the liquid scintillator. The two dashed lines represent the fine-tuning dates of the simulation. The solid line marks the transition between phases II and III and the corresponding change of parameter set for the simulation fine-tuning. The very small discontinuity of the CCs at this point validates the fact that two fine-tunings of the simulation are sufficient for a precise control of the detector response over the whole experiment. \textbf{b.} Residual fluctuations over time of the position of the 2.2~MeV peak from capture of cosmic neutrons on hydrogen (n-H peak), obtained after application of the energy reconstruction with the evolving CCs. The $1\sigma$ uncertainty of each point is taken from the fit of the experimental peaks with a crystal ball function. The RMS of the relative deviations to the mean value (0.25\%) is used as an estimate of the systematic uncertainty on the time stability of the energy reconstruction. \textbf{c.} Time evolution of the energy resolution of the n-H peak, setting the scale for the average resolution in the full target volume and its time evolution due to the slow decrease of collected light. The fine-tuning of the MC accounts for the different mean resolutions in phase-II and III.}
    \label{fig:nH}
\end{extdatafigure*}

\begin{extdatafigure*}[!ht]
    \centering
    \includegraphics[width=\linewidth]{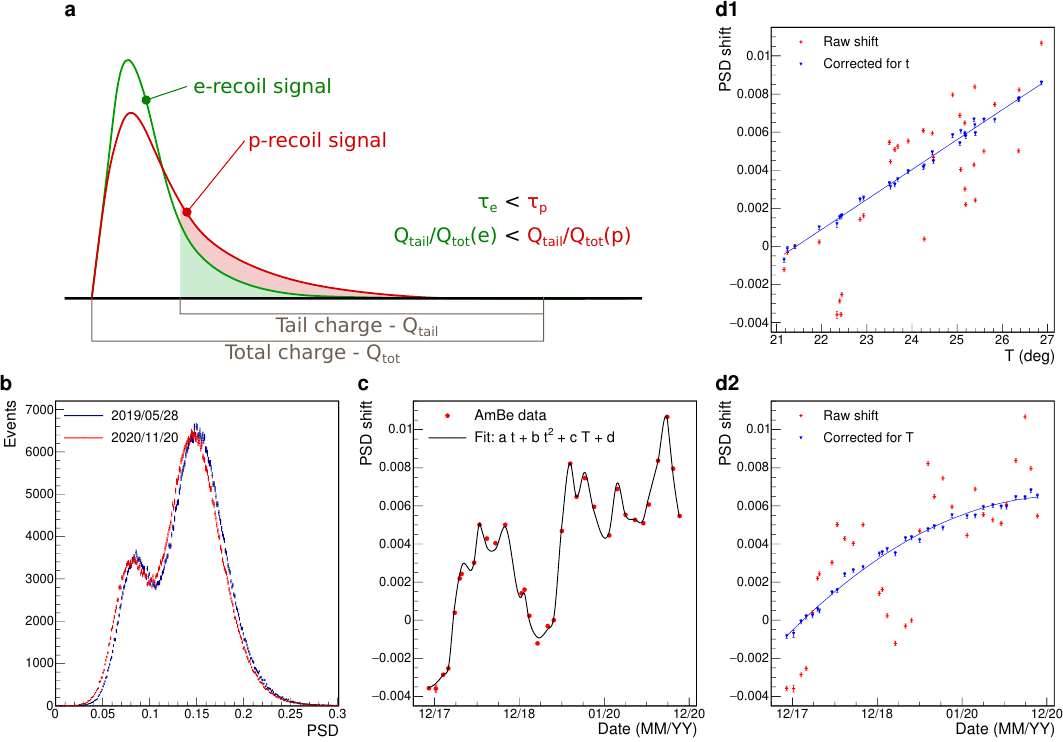}
    \caption{\textbf{Stability of the Pulse Shape Discrimination variable.} \textbf{a.} Illustration of the Pulse Shape Discrimination method, that separates electron recoils and proton recoils based on the shape of the collected light pulse. The PSD variable is defined as $Q_\mathrm{tail}/Q_\mathrm{tot}$. \textbf{b.} PSD distributions of Am-Be events from a run in May 2019 (reference date for the correction) and another in November 2020 (end of the data taking), showing the distortion over time (mostly a shift of the distribution). All plots are for cell 4 and reconstructed energies between 1.625~MeV and 4.125~MeV. \textbf{c.} Evolution of the  shift required to align the PSD distribution of any given date to the reference of May 2019. The time-temperature polynomial fit is displayed. \textbf{d1.} Evolution of the shift with temperature: once the time dependence is corrected, the evolution follows a first-order polynomial. \textbf{d2.} Evolution of the shift over time: once the temperature dependence is corrected, the evolution follows a second-order polynomial. All displayed uncertainties are statistical at $1\sigma$ CL.}
    \label{fig:PSD}
\end{extdatafigure*}

\begin{extdatafigure*}
    \centering
    \includegraphics[width=\linewidth]{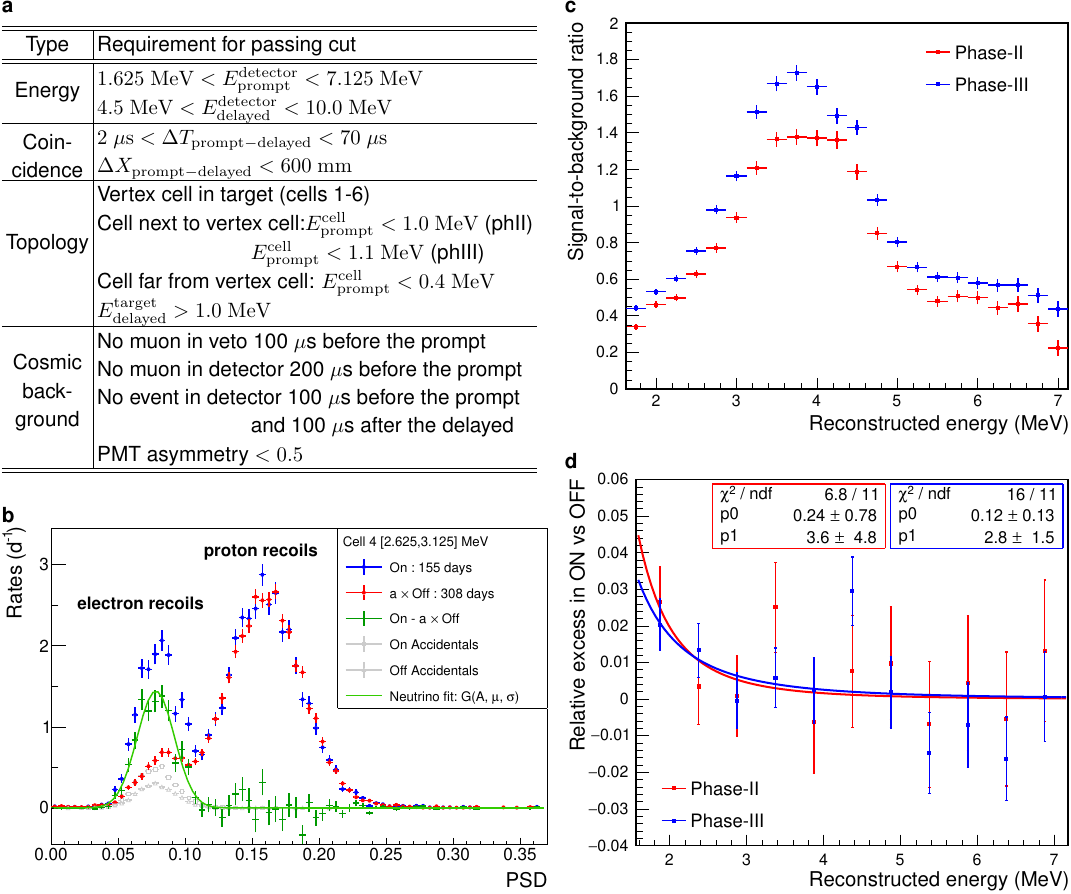}
    \caption{\textbf{IBD signal and backgrounds.} \textbf{a.} Selection cuts for IBD candidates (see text for details). The topology cut on $E^\mathrm{cell}_\mathrm{prompt}$ was loosened for phase-III because of detector ageing. \textbf{b.} Illustration of the extraction of the IBD signal using event distributions based on the PSD variable$Q_\mathrm{tail}/Q_\mathrm{tot}$ (after application of the selection cuts) for phase-III. The two populations ($e$-recoils and $p$-recoils) are well separated. The pair rate in ON (blue) is the sum of: i) correlated background pairs (red, rescaled from OFF data), ii) accidental pairs (grey squares), iii) the IBD signal (green, modelled by a Gaussian whose integral gives the IBD rate in this bin). The scaling factor $a$ on the OFF distribution depends on environmental parameters (e.g. atmospheric pressure) and ON/OFF relative running time; it is treated as a free parameter in the fit. \textbf{c.} Signal-to-background ratios obtained at the target level (combining the 6 cells), for phase-II and phase-III. Higher reactor power in phase-III led to more IBD signal. Only events with PSD values in the signal range, defined as lower than the ``mean position + 2.5 sigmas'' of the electronic recoils, are used to produce this plot.  \textbf{d.} Search for reactor-related background events in the proton-recoil region from the ON $-\,a\times$OFF distributions at the target level. A low-energy excess is found and fitted by a power law $f(E) = p_0 E^{-p_1}$. Here the $a$ parameter is not fitted but fixed from the measured sensitivity of the cosmic background to atmospheric pressure and the mean pressure difference between the ON and OFF periods; the consistency with zero at high energy is a good validationof the extraction method (fitted and computed values of $a$ are in agreement). All displayed uncertainties are statistical at $1\sigma$ CL.}
    \label{fig:SB}
\end{extdatafigure*}

\begin{extdatafigure*}
    \centering
    \includegraphics[width=0.75\linewidth]{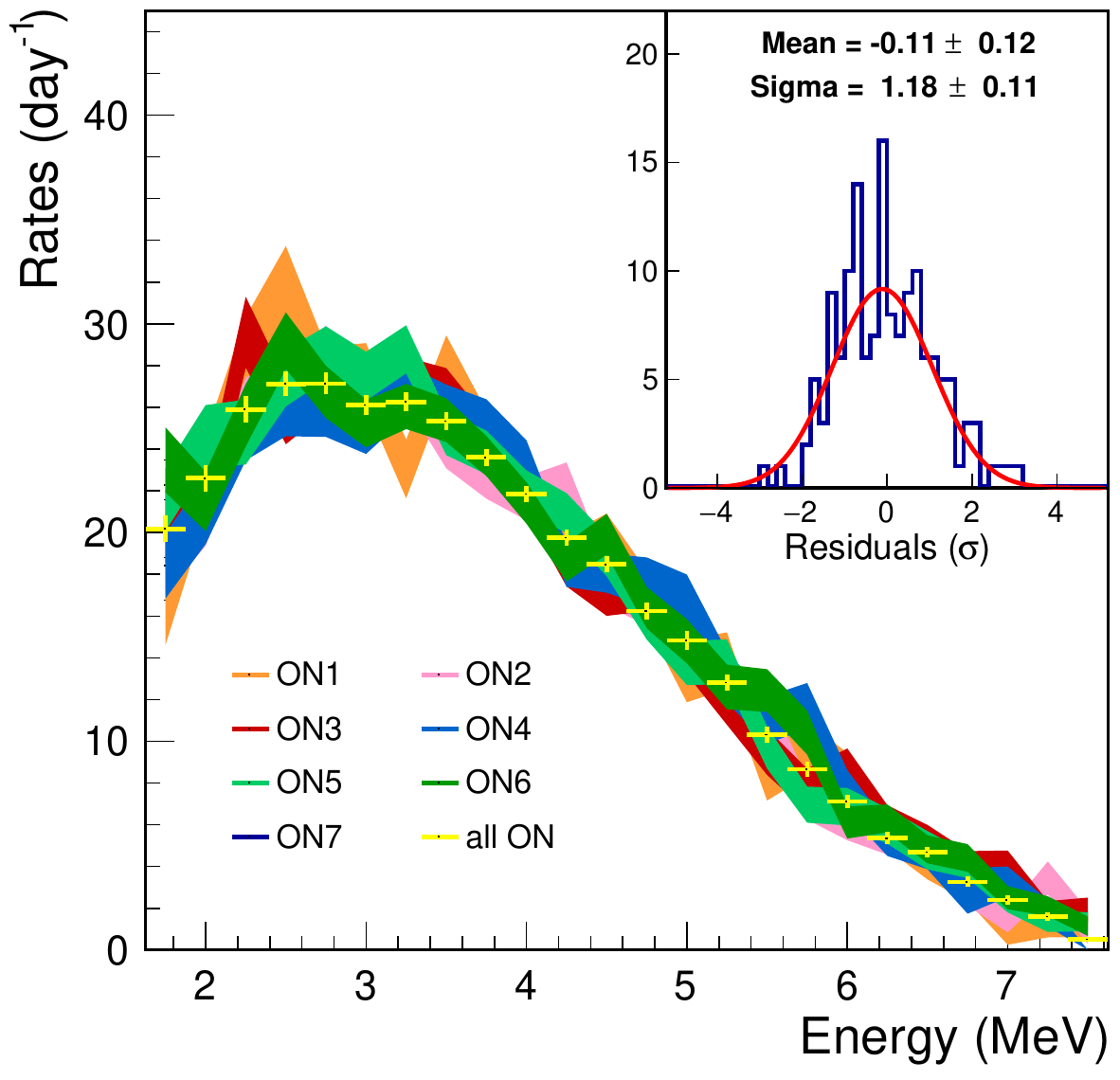}
    \caption{\textbf{Time stability of the extracted antineutrino spectra}. IBD spectra are extracted from each of the 7 acquired reactor cycles using adjacent OFF data for background subtraction. The colored bands illustrate the $1\sigma$ statistical uncertainties. Inset: for each individual ON spectrum the residuals with respect to the average of all other spectra are computed. The distribution of all these residuals is found compatible with a normal distribution.}
    \label{fig:ONcomp}
\end{extdatafigure*}

\begin{extdatafigure*}[t]
    \centering        
    \includegraphics[width=\linewidth]{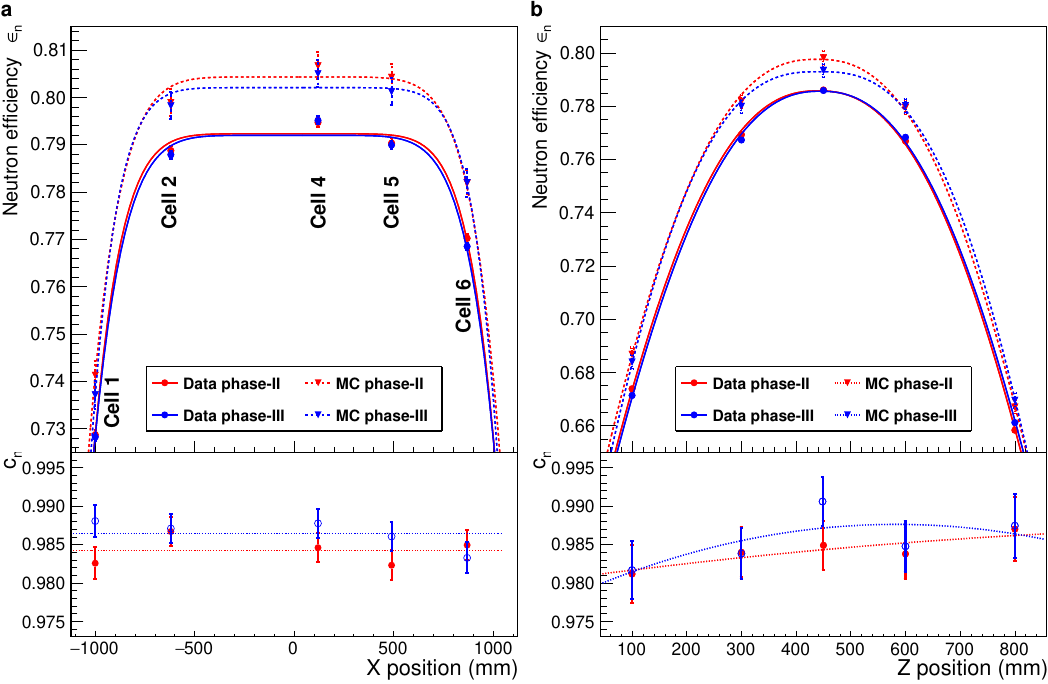}
    \caption{\textbf{Neutron efficiency.} Top panels: neutron efficiency $\varepsilon_\mathrm{n}$ obtained with the Am-Be source as function of (a) horizontal (averaged over all measured heights) and (b) vertical (averaged over all measured cells) position for phase-II (red) and phase-III (blue). Measurements (circles) are performed in cells 1, 2, 4, 5 and 6 at heights $Z=$10~cm, 30~cm, 45~cm, 60~cm and 80~cm. Simulated efficiencies obtained for the same positions are shown as triangles. A 3D spatial model is fitted and gives a continuous description of the efficiency in the detector: $\varepsilon_\mathrm{n}^\mathrm{data}(X,Y,Z)$ (solid lines) and $\varepsilon_\mathrm{n}^\mathrm{MC}(X,Y,Z)$ (dashed lines). Bottom panels: the coefficient $c_\mathrm{n} = \varepsilon_\mathrm{n}^\mathrm{data} / \varepsilon_\mathrm{n}^\mathrm{MC}$ is used to correct for efficiency biases due to neutron simulation imperfections. It is consistent with a constant function of $X$ and a 2nd order polynomial in $Z$; the difference in the integrated $\overline{c_\mathrm{n}}$ coming from the choice of a $(X,Y,Z)$ model for $c_\mathrm{n}$ is considered as systematic uncertainty. All displayed uncertainties are statistical at $1\sigma$ CL.}
    \label{fig:neuteff}
\end{extdatafigure*}

\begin{extdatafigure*}
    \centering
    \includegraphics[width=\linewidth]{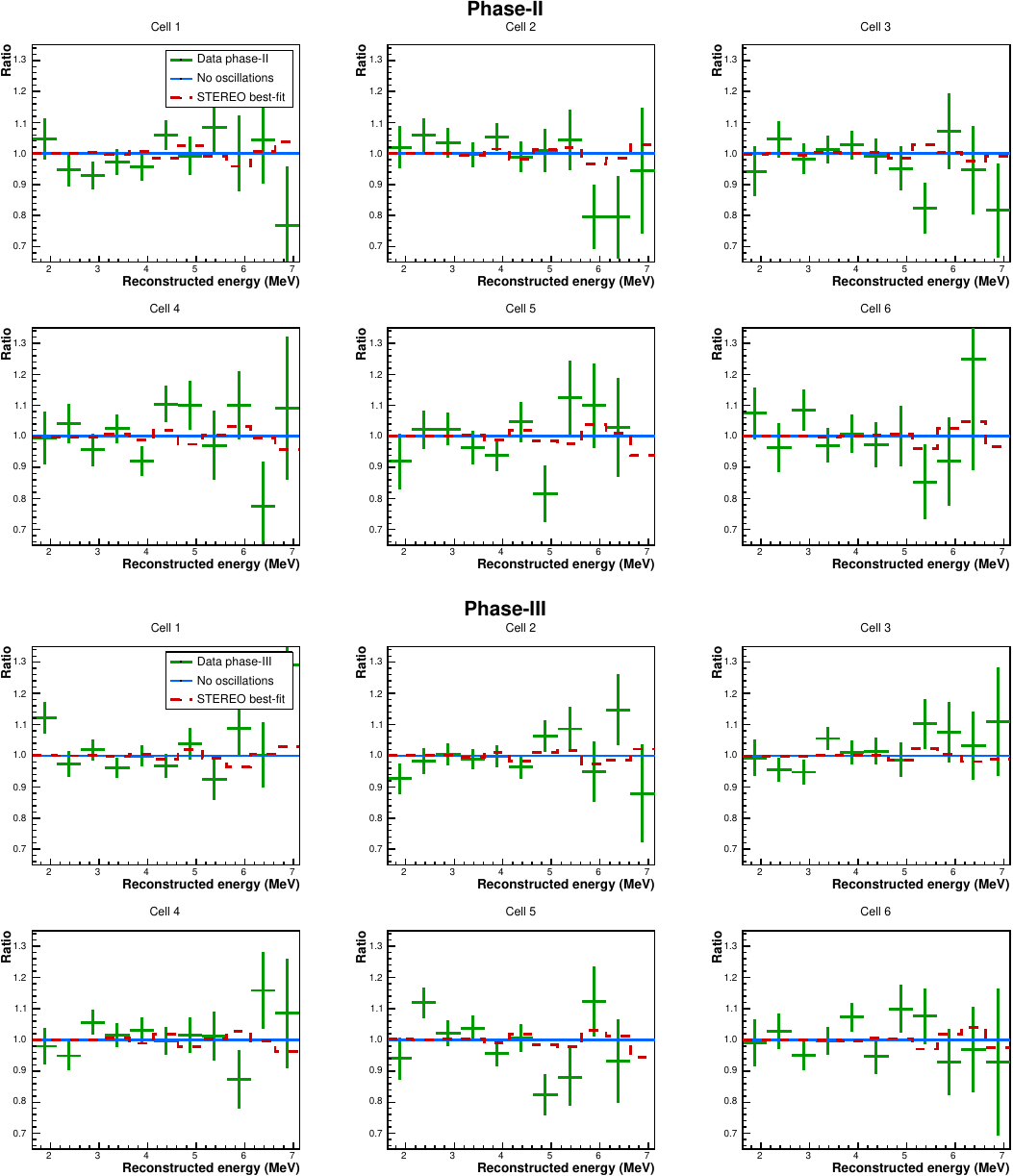}
    \caption{\textbf{Antineutrino spectra in each cell relative to the no-oscillation model.} Antineutrino spectra are displayed as a ratio to the adjusted no-oscillation prediction $\hat{\phi}_i M_{p,l,i}(0,0;\hat{\alpha}^j)$ for cell $l \in \{1, \dots, 6\}$ and phase $p \in \{\mathrm{II}, \mathrm{III}\}$. The best-fit $\hat{\phi}_i$ parameters, common to all cells and phases, absorb an overall spectrum shape so that only relative distortions between cells remain. No significant deviation is found between data and the no-oscillation prediction. For illustration we displayed also the spectra induced in each cell by the best-fit sterile oscillation parameters (dashed red line). All displayed uncertainties are statistical at $1\sigma$ CL.}
    \label{fig:cell_spectra}
\end{extdatafigure*}

\begin{extdatafigure*}
    \centering
    \includegraphics[width=0.70\linewidth]{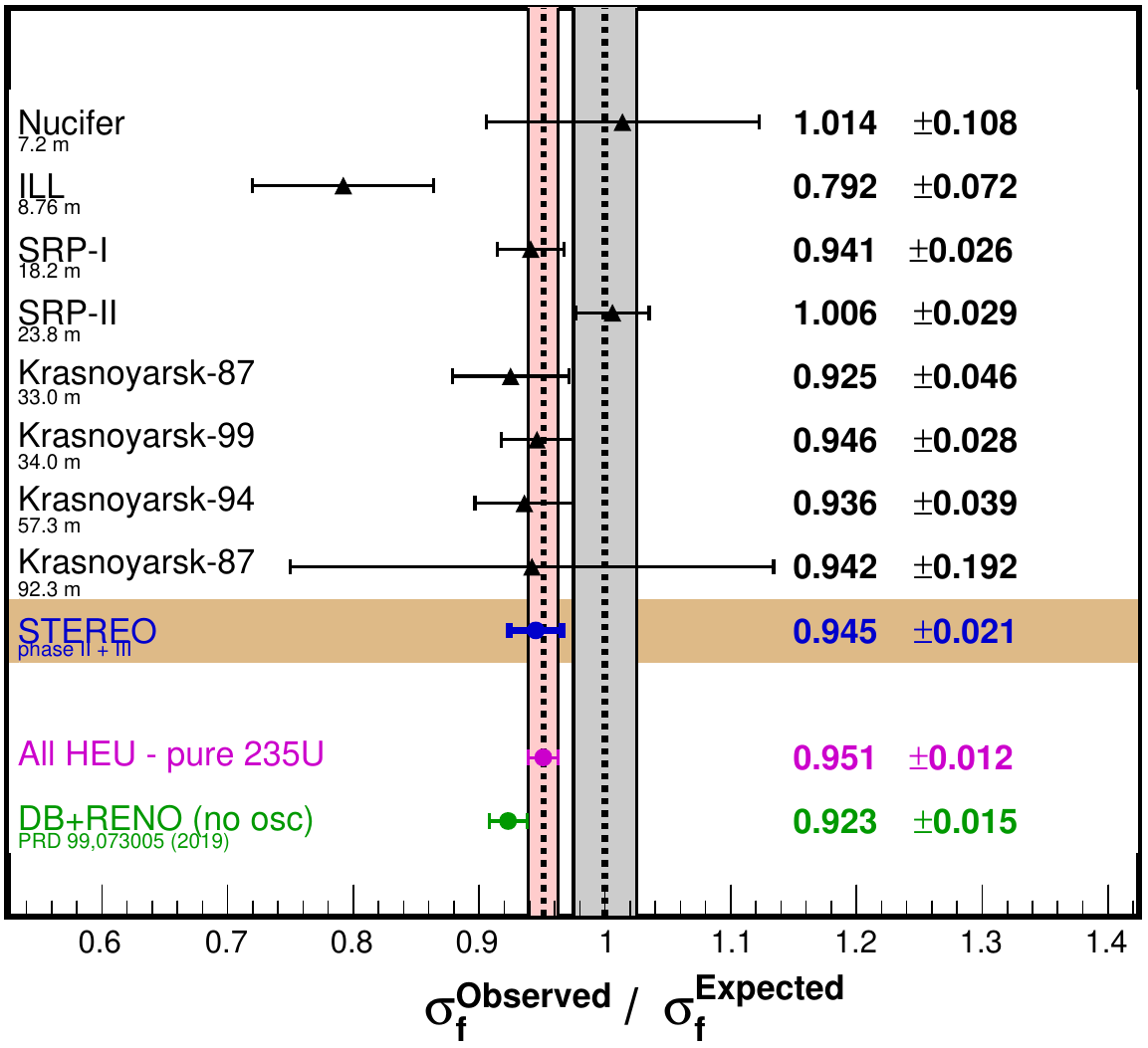}
    \caption{\textbf{Antineutrino yield of $^{235}$U fission.} Overview of the measured antineutrino flux from pure fission of $^{235}$U (highly enriched nuclear fuel) relative to the HM model. For a direct comparison of data from different detectors with different thresholds and resolutions, the quantity of interest is the ratio of the measured to expected cross sections per fission, $\sigma_f$, defined as the integral of antineutrino spectrum multiplied by the IBD cross section. The measurement by \textsc{Stereo} (0.945 $\pm$ 0.021) is the most accurate to date and found to be in excellent agreement with the previous world average of 0.954 $\pm$ 0.014 taken from \cite{Gariazzo:2017fdh}. For comparison, we also display the measurement from Daya Bay and RENO with commercial reactors (lowly enriched nuclear fuel, green) although it relies on reactor evolution simulations to separate the contribution of $^{235}$U from other isotopes. The size of the error bars corresponds to the total uncertainty of the respective measurement. The uncertainty of the HM model, common to all measurements and not included in the error bars, is illustrated by the width of the grey band while the purple point and band represent the central value and the uncertainty of the new world average, after including our result. All displayed uncertainties are at $1\sigma$ CL.}
    \label{fig:lep_plot}
\end{extdatafigure*}

\begin{extdatatable*}[t]
    \centering
    \includegraphics[width=0.9\linewidth]{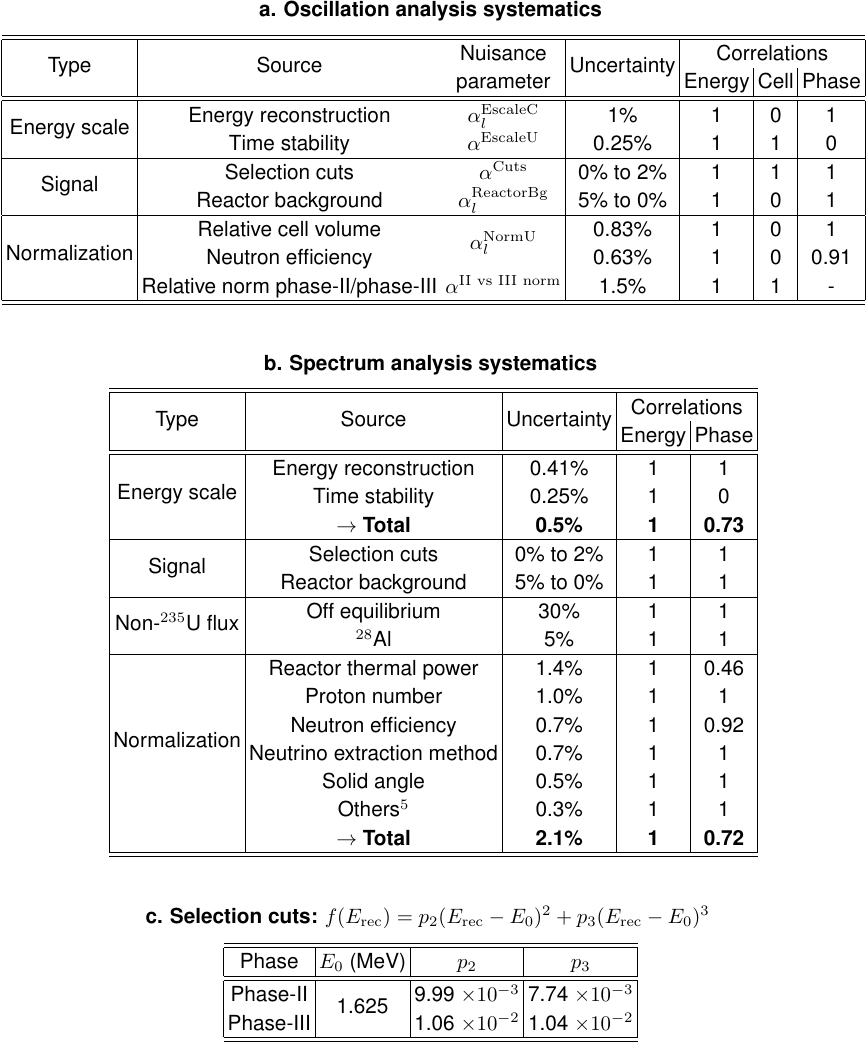}
    \caption{\textbf{Systematics uncertainties.}
    \textbf{a.} The oscillation analysis studies relative distortions between spectra in each cell, so systematics related to the overall spectrum shape (\textit{e.g.} non-$^{235}$U flux) or to a common normalization of neutrino rates in all cells (\textit{e.g.} solid angle) are not included. Specifically, only part of the neutron efficiency uncertainty is relevant to the oscillation analysis. Nuisance parameters refer to the $\chi^2$ expression of Eq.~(\ref{eqn:chi2}).
    \textbf{b.} The spectrum analysis is based on the target spectrum (sum over all cells): uncertainties that are uncorrelated between cells are thus reduced at the target level (\textit{e.g.} energy reconstruction). 
    In \textbf{a.} and \textbf{b.} the energy dependence for uncertainty from selection cuts is given by the $f(E_\mathrm{rec})$ function defined in \textbf{c.}; for reactor background, it corresponds to the power law fitting the observed low-energy excess divided by the signal-to-background ratio (both displayed in Extended Data Fig.~\ref{fig:SB}).
    \textbf{c.} Parameters describing the energy dependence of the uncertainty associated to selection cuts, for each phase. Their extraction is detailed in \cite{STEREO:2020hup}.}
    \label{tab:syst}
\end{extdatatable*}

\end{appendices}
\end{document}